
\input phyzzx

\Pubnum{ \vbox{ \hbox{R/95/26} }}

\pubtype{}
\date{August 1995}

\titlepage

\title{Supersymmetric Toda field theories}

\author{G. Papadopoulos}
\address{D.A.M.T.P.\break University of Cambridge\break Silver Street\break
Cambridge CB3 9EW}

\abstract{We present new supersymmetric extensions of Conformal Toda and
$A^{(1)}_N$ Affine Toda field theories. These new theories are constructed
using
methods similar to those that have been developed to find supersymmetric
extensions of two-dimensional bosonic sigma models with a scalar potential. In
particular, we show that the Conformal Toda field theory admits a
(1,1)-supersymmetric extension, and the $A^{(1)}_N$ Affine Toda field admits a
(1,0)-supersymmetric extension.}
\endpage



\def\G {\bf {g}}
\def\H {\bf {h}}

\def\fff{\vrule width0.5pt height5pt depth1pt}
\def\pp{{{ =\hskip-3.75pt{\fff}}\hskip3.75pt }}

\def\a{\alpha}


\REF\asea{ M.A. Olshanetsky, Commun. Math. Phys. {\bf 88} (1983) 63.}
\REF\aseb{J. Evans and T. Hollowood, Nucl. Phys. {\bf B352} (1991) 723.}
\REF\asec{A. Gualzetti, S. Penati and D. Zanon, Nucl. Phys. {\bf B398} (1993)
622.}
\REF\lsl{A.N. Leznov, M.V. Saveliev and D.A. Leites, Phys. Lett. {\bf 96B} 97.}
\REF\ap{A.M. Polyakov, Phys. Lett. {\bf 103B} (1981) 211.}
\REF\jfa{ J.F. Arvis, Nucl. Phys. {\bf B212} (1983) 151; {\it ibid} {\bf B218}
(1983) 309.}
\REF\edh{E. D'Hoker, Phys. Rev. {\bf D28} (1983) 1346.}
\REF\cg{T. Curtright and G. Ghandour, Phys. Lett. {\bf 136B} (1984) 50.}
\REF\ob{O. Babelon, Phys. Lett. {\bf 141B} (1984) 353; Nucl. Phys. {\bf B258}
(1985) 680.}
\REF\bl{O.Babelon and F. Langouche, Nucl. Phys. {\bf B290} (1987) 603.}
\REF\df {P. Di Vecchia and S. Ferrara, Nucl. Phys. {\bf B130} (1977) 93.}
\REF\fgs{S. Ferrara, L. Girardello and S. Sciuto, Phys. Lett. {\bf 76B} (1978)
303.}
\REF\jh{J. Hruby, Nucl. Phys. {\bf B131} (1977) 275.}
\REF\ms{T. Marincucci and S. Sciuto, Nucl. Phys. {\bf B156} (1979) 144.}
\REF\sw{R. Shankar and E. Witten, Phys. Rev. {\bf D17} (1978) 2134.}
\REF\ku{K. Kobayashi, T. Uematsu and Y. Yu, Nucl. Phys. {\bf 397} (1993) 283.}
\REF\lag{L. Alvarez-Gaum{\' e} and D.Z. Freedman, Commun. Math. Phys.
{\bf 91} (1983), 87.}
\REF\gpa{ C.M. Hull, G. Papadopoulos and P.K. Townsend, Phys. Lett.
{\bf 319B} (1994) 291.}
\REF\gpb{G. Papadopoulos and P.K. Townsend, Class. Quantum Grav. {\bf 11}
(1994)
515;{\it ibid} {\bf 11} (1994) 2163.}
\REF\gpc{G. Papadopoulos and P.K. Townsend, Nucl. Phys. {\bf B444} (1995) 245.
}
\REF\tjh {T.J. Hollowood, Nucl. Phys. {\bf B384} (1992) 523.}
\REF\hou{D.I. Olive, N. Turok and J.W.R. Underwood, Nucl. Phys. {\bf B401}
(1993)
663.}

\sequentialequations


\chapter{Introduction}

It has been known for sometime that the Conformal Toda and Affine Toda field
theories admit supersymmetric extensions.  These extensions were based on the
theory of superalgebras and the emphasis was to construct {\sl integrable}
(1,1)-
and (2,2)-supersymmetric extensions  of Conformal Toda and Affine Toda field
theories [\asea-\asec].  This led to theories that, apart from the
supersymmetric
extensions of Liouville theory [\lsl-\bl], and the sinh- and sine-Gordon models
[\df-\ku],  either their supersymmetry was brocken explicitly at the level of
their
classical action or their energy was unbounded from below. (The inner product
that
appears in their kinetic term is indefinite).

The Conformal and Affine Toda theories are field theories of scalar fields with
a
scalar potential interaction and are examples of sigma models with a scalar
potential and flat space as a target manifold.   Recently, there has been a lot
of progress towards classifying all (p,q)-supersymmetric two-dimensional sigma
models with a scalar potential [\gpa, \gpb].  These results were used in ref.
[\gpc] to find a supersymmetric extension of the WZW model with a scalar
potential
and show that it has a number of novel properties like a multi-vacuum
structure,
solitons that saturate a Bogomol'nyi bound and supersymmetry breaking for
certain
values of its coupling constants.  Since Conformal and Affine Toda field
theories
are special cases of two-dimensional sigma models, it seems that the methods
applicable for constructing supersymmetric extensions of sigma models with
scalar
potentials can be used to construct supersymmetric extensions of the Conformal
and
Affine Toda field theories.

The main purpose of this letter is to present supersymmetric extensions of  the
Conformal and $A^{(1)}_N$ Affine Toda field theories using the general methods
applied to construct supersymmetric extensions of bosonic sigma models with a
scalar potential.  For this, we will start from the bosonic action of Conformal
and
$A^{(1)}_N$ Affine Toda field theories and  we will then add the kinetic terms
of
the fermions and the interaction terms of the fermions with the scalar fields.
These supersymmetric  Conformal and $A^{(1)}_N$ Affine Toda field theories will
have the following properties: (i) the classical vacuum of the Conformal and
Affine Toda field theories will be  supersymmetric\foot{We have assumed that
the
coupling constant is real.} and (ii) the mass matrix of the fermions will be
well
defined (finite) at this vacuum. The energy functional of all these theories is
positive semi-definite. In the case of Conformal Toda field theory, it is
straightforward to find a `minimal' (1,0)-supersymmetric extension. In fact,
this (1,0)-supersymmetric Conformal Toda field theory can be further extended
to a
(1,1)-supersymmetric one. The (bosonic) Conformal Toda field theory has a
unique
vacuum at infinity.  It turns out that this vacuum is a supersymmetric and the
fermion mass matrix evaluated at this vacuum vanishes.  In the case of
$A^{(1)}_N$ Affine Toda field theory, we will show that the theory admits a
(1,0)-supersymmetric extension for which the properties (i) and (ii) above
are satisfied and that this (1,0)-supersymmetric extension for,
$N\geq 2$, is not unique.   Finally, we will compute the
supersymmetry charges of (1,1)-supersymmetric Conformal Toda and
(1,0)-supersymmetric $A^{(1)}_N$ Affine Toda field theories and explicitly give
the
fermion mass matrix of the supersymmetric $A^{(1)}_2$ Affine Toda field theory.

 The material of this letter is organised as follows:  In section 2,  the
action
and supersymmetry transformation laws of the fields of the (1,0)- and
(1,1)-supersymmetric sigma models with scalar potential will be presented.  In
section 3, a (1,1)-supersymmetric extension of the Conformal Toda field theory
will be given.  In section 4, two different (1,0)-supersymmetric extensions of
the
$A_N^{(1)}$ Affine Toda field theory will be presented and in section 5, we
will
give our conclusions.


\chapter{(1,0)-and (1,1)-supersymmetric massive sigma models}

The (p,q) supersymmetry algebra in two dimensions is
$$
\{Q_+^I, Q_+^J\}=2 \delta^{IJ}(E+P)\ ,\qquad  \{Q_-^{I'}, Q_-^{J'}\}=2
\delta^{I'J'} (E-P)\ ,\qquad \{Q_+^I, Q_-^{I'}\}= Z^{II'}\ ,
\eqn\inone
$$
where $\{Q_+^I; I=1,\dots, p\}$ are the `left' supersymmetry charges,
$\{Q_+^{I'}; I'=1,\dots, q\}$ are the `right' supersymmetry charges, $Z^{II'}$
are central charges, $E$ is the energy and $P$ is the momentum generators.
We will be primarily concerned here with the (1,0) and (1,1) supersymmetry
algebras.  The (1,0) supersymmetry algebra has one `left' supersymmetry charge
$Q_+$ and no central charges.  The (1,1) supersymmetry algebra has one `left'
supersymmetry charge $Q_+$, a `right' supersymmetry charge $Q_-$ and it may
also have one central charge $Z$. For a bosonic sigma model to admit a
(p,q)-supersymmetric extension several conditions of the couplings of the
theory
must be imposed.  These conditions have a geometric interpretation in terms of
the
geometry of the sigma model target space and for the case of sigma model with a
scalar potential were given in refs. [\lag, \gpa, \gpb].  Let $g$ be a metric
and
$b$ be a locally defined two-form on a manifold ${\cal M}$.  In addition, let
$A$
be a connection of a vector bundle ${\cal E}$ over ${\cal M}$ and
$k$ be a fibre metric of ${\cal E}$. The action of most general
(1,0)-supersymmetric sigma model with scalar potential
is\foot{This theory admits an off-shell (1,0) superspace formulation and this
action is derived after elimination of the auxiliary fields [\gpa].}
$$
\eqalign{
I =\int\! d^2 x\big\{ &\partial_\pp\phi^i\partial_=\phi^j (g_{ij}+b_{ij})
+ ig_{ij}\lambda_+^i\nabla_=^{(+)}\lambda_+^j -
ik_{ab}\psi_-^a\nabla_\pp\psi_-^b\cr
& -{1\over2}\psi_-^a\psi_-^b\lambda_+^i\lambda_+^j F(A)_{ijab}
 +m\nabla_i s_a \lambda_+^i\psi_-^a  -V(\phi) \big\}\ ,}
\eqn\intwo
$$
where $\phi$ is the sigma model field, $\lambda_+$ and $\psi_-$ are
real chiral fermions, $\{x^\pp, x^=\}$ are the light-cone co-ordinates of the
two-dimensional Minkowski space-time,
$$
\nabla_\pp\psi^a_-=\partial_\pp \psi_-^a +\partial_\pp\phi^i
A_i{}^a{}_b\psi^b_-\ ,
\eqn\inthree
$$
$F(A)$ is the curvature of $A$, $s$ is a section of the vector bundle ${\cal
E}$,
$$
V(\phi)={m^2\over 4} k^{ab} s_a(\phi) s_b(\phi)\ ,
\eqn\infour
$$
is the scalar potential and $m$ is a parameter with dimension that of a mass.
Finally, the connections of the covariant derivatives $\nabla^{(\pm)}$ are
$$
\Gamma^{(\pm)}{}^i{}_{jk}=\{{}^i_{jk}\}\pm H^i{}_{jk}
\eqn\infive
$$
where $H={3\over2}db$.
{}From the action \intwo, the fermion mass matrix is
$$
M_{ia}=m \nabla_is_a\ .
\eqn\infoura
$$
It is important to note that the range of the sigma model manifold indices
$i,j$
is different from that of the vector bundle indices $a,b$.
The action \intwo\ is invariant under the (1,0) supersymmetry transformations
$$\eqalign{
\delta_{\epsilon}\phi^i&=-{i\over2}\epsilon_- \lambda_+^i
\cr
\delta_{\epsilon}\lambda_+^i&={1\over2} \epsilon_- \partial_\pp\phi^i
\cr
\delta_{\epsilon}\psi_-^a&= {i\over2}\epsilon_- A_i{}^a{}_b \lambda_+^i
\psi^b_-+{i\over4}
\epsilon_- m s^a\ , }
\eqn\infivea
$$
where $\epsilon_-$ is the parameter of the transformations.
The (1,0)-supersymmetry charge is
$$
S_+=\int dx [g_{ij}\partial_\pp\phi^i \lambda^j_+-i{1\over3}H_{ijk} \lambda^i_+
\lambda^j_+ \lambda^k_+-{i\over 2} m s_a
\psi^a_-]\ .
\eqn\insix
$$

To summarise, a bosonic sigma model with a scalar potential admits a
(1,0)-super\-sym\-metric extension if and only if its scalar potential can be
written as the length of a section of a vector bundle over the sigma model
manifold.  The {\sl supersymmetric} vacua of the theory are simply the {\sl
zeroes} of this section. Note that there might be different ways, up to gauge
equivalences, to write the scalar potential of a bosonic sigma model as the
length
of a section of a vector bundle over the sigma model manifold. Different
choices
may lead to different (1,0)-supersymmetric extensions of the same bosonic sigma
model.

Next we will describe the most general (1,1)-supersymmetric sigma models with
scalar potential [\gpa]. These models are special cases of (1,0)-supersymmetric
sigma models with scalar potential and are those for which the vector
bundle ${\cal E}$ is isomorphic to the tangent bundle of ${\cal M}$, the
connection $A$ is the spin connection of $\Gamma^{(-)}$ and the section
$s$ is
$$
s_i=(u -X)_i
\eqn\inseven
$$
where $X$ is a Killing vector field of ${\cal M}$ that leaves $H$ invariant,
the one-form
$u$ is defined from the relation
$$
X^kH_{kij}=\partial_{[i} u_{j]}\ ,
\eqn\ineight
$$
and $X^i u_i=0$.
The (0,1) supersymmetry transformations are
$$
\eqalign{
\delta_\zeta\phi^i&=\zeta_+ \psi^i_-
\cr
\delta_\zeta\lambda^i_+&= m {1\over2}\zeta_+
(u+X)^i+ \zeta_+\Gamma^{(-)}{}^i{}_{jk}
\lambda_+^j \psi_-^k
\cr
\delta_\zeta\psi_-^i&=-i \zeta_+\partial_=\phi^i\ ,}
\eqn\ineighta
$$
where $\zeta_+$ is the parameter.
The (0,1)-supersymmetry charge is
$$
S_-=\int dx [ i g_{ij} \partial_=\phi^i \psi^j_-
 + {1\over 3} \psi^i_- \psi^j_- \psi^k_-
H_{ijk} + {m\over 2} (X+u)_i \lambda_+^i]\ .
\eqn\innine
$$
In the models of interest here, the torsion $H$ and the Killing vector field
$X$
are zero in which case $u=df$ where $f$ is a  function of the sigma model
manifold ${\cal M}$.


\chapter{A Supersymmetric Conformal Toda field theory}

Let $\G$ be a simple Lie algebra and  $\H$ be a Cartan subalgebra of $\G$
($\rm{dim}\H\equiv \rm{rank} \G=r$).The action of the bosonic Conformal Toda
field
theory is
$$
I= \int d^2x \big(<\partial_\pp\phi, \partial_=\phi> -{m^2\over4} \sum^r_{i=1}
\mu^i e^{\a_i (\phi)}\big)\ ,
\eqn\tone
$$
where $\phi$ are the fields that are maps from the two-dimensional space-time
into the Cartan subalgebra $\H$,  $<\cdot,\cdot>$ is the bi-invariant inner
product of $\G$ restricted on $\H$, $\{\a_i; i=1, \dots, \rm{rank}\G\}$ are the
simple roots  of $\G$, $\a_i(\phi)=\a_{in}\phi^n$ and
$\{\mu^i; i=1, \dots,r\}$ are {\sl positive} real constants\foot{The index $n$
is
associated with an orthonormal basis of the simple Lie algebra $\G$. }.  It is
clear that this theory is a sigma model without Wess-Zumino term ($b=0$) and
flat
target space.  The scalar potential $V$ of the theory is given by
$$
V(\phi)={m^2\over4} \sum^r_{i=1} \mu^i e^{\a_i (\phi)}\ .
\eqn\taone
$$

To construct a (1,0)-supersymmetric extension of this theory, we first
introduce
chiral fermions
$\lambda_+$ and
$\psi_-$ and then choose
$$
\{s_n; n=1,\dots, r\}=\{\sqrt {\mu^i}  e^{{1\over2}\a_i (\phi)}; i=1,
\dots, r\}\ .
\eqn\ttwo
$$
For this choice of $s$, the fibre metric $k$ is $k=\rm{diag}(1,\dots,1)$.
The action of the supersymmetric Toda field theory is
$$\eqalign{
I=&\int d^2x \big(<\partial_\pp\phi, \partial_=\phi> + i<\lambda_+,
\partial_=\lambda_+>-i <\psi_-,\partial_\pp \psi_->
\cr &
+{m\over2} \sum^r_{i=1} \sqrt{\mu^i} \a_i(\lambda_+)
 e^{{1\over2}\a_i (\phi)} \psi_-^i -{m^2\over4} \sum^r_{i=1} \mu^i e^{\a_i
(\phi)}
\big)\ .}
\eqn\tthree
$$
The fermion mass matrix is
$$
M_{in}={m\over2}\a_{in}\sqrt {\mu^i}  e^{{1\over2}\a_i (\phi)}\ .
\eqn\tfour
$$
The vacuum of the theory is in the limit $\a_i(\phi)\rightarrow -\infty$ and in
this limit the scalar potential,$V$, vanishes. This vacuum is also
supersymmetric
because the section $s$ of eqn \ttwo vanishes as $\a_i(\phi)\rightarrow
-\infty$.
Moreover the bosons and the fermions of the theory are massless in this vacuum.

The above (1,0)-supersymmetric Conformal Toda field theory can be further
extended to a (1,1) supersymmetric one.  For this, we observe that all the
conditions mentioned in the previous section for a (1,0)-supersymmetric sigma
model
to be a (1,1)-supersymmetric one are satisfied.  In particular, we note that
$$
\{s_n; n=1,\dots, r\}=\{\partial_i f(\phi); i=1,
\dots, r\}\ ,
\eqn\tafour
$$
where
$$
f=2\sum^r_{i=1} \sqrt {\mu^i}  e^{{1\over2}\a_i (\phi)}\ ,
\eqn\tbfour
$$
$\partial_i=\partial/\partial\phi^i$ and $\phi^i=\a_i (\phi)$.

The (1,0) and (0,1) supersymmetry charges are given as follows:
$$
S_+=\int dx [<\partial_\pp\phi^i,
\lambda_+> -i{m\over 2} \sum^r_{i=1} \sqrt {\mu^i}  e^{{1\over2}\a_i(\phi)}
\psi^i_-]\ .
\eqn\tfive
$$
and
$$
S_-=\int dx[i<\partial_=\phi, \psi_->+ {m\over 2}
\sqrt {\mu^i}  e^{{1\over2}\a_i (\phi)}\delta_{in} \lambda^n_+] \ ,
\eqn\tseven
$$
correspondingly. It can be verified by direct computation that the Poison
bracket algebra of these charges is isomorphic to the (1,1) supersymmetry
algebra \inone. Finally, note that the action, \tthree,
of the (1,1)-supersymmetric Conformal Toda field theory can be written in terms
of
(1,1) off-shell superfields.


\chapter{Supersymmetric Affine Toda}

The action of bosonic Affine Toda field theory is\foot{The Lagrangian of the
Affine
Toda field theory is usually  written without the term involving the Coexeter
number $h$. This term is not necessary in the bosonic theory since it does
not contribute to the field equations.  However in the supersymmetric theory,
this
is no longer the case because the absence of this term from the action will
lead
to a theory with spontaneously brocken supersymmetry.}
$$
I=\int d^2x\, \big[<\partial_\pp\phi,\partial_=\phi>-{m^2\over
4\beta^2}\big(\sum^r_{i=0} n_i e^{\beta a_i(\phi)}- h\big)\big]
\eqn\atone
$$
where $\phi$ is a map from the two-dimensional Minkowski space-time into a
Cartan subalgebra $\H$ of a simple group $\G$, $r=\rm {rank} \G$, $a_i=\a_i$
for $i\geq 1$ and
$a_0=-\psi$, $\psi$ is the highest root of $\G$, the integers $n_i$ for $i\geq
1$ are defined by the equation
$$
\psi=\sum^r_{i=1} n_i \alpha_i\ ,
\eqn\atonea
$$
$n_0=1$ and $h$ is the Coexeter number, i.e. $h=\sum^r_{i=0} n_i$.  The
parameters
$m$ and $\beta$ are the couplings constants of the theory.

Next consider the special case of $A_N^{(1)}$ Affine Toda field theory.  In
this
case $r=N$, $n_i=1$ and $h=N+1$. The scalar potential of the $A_N^{(1)}$ Affine
Toda field theory is
$$
V_N(\phi^1, \dots, \phi^N)={m^2\over 4\beta^2}[\sum^N_{i=1}
e^{\beta \phi^i}+ e^{-\beta{\sum^N_{i=1}\phi^i}}- (N+1)]\ ,
\eqn\atonea
$$
where $\phi^i=\a^i(\phi)$.
To show that the $A_N^{(1)}$ Affine Toda field theory admits a
(1,0)-super\-sym\-metric extension we have to write the scalar potential of the
theory as the sum of squares in such a way that the following two properties
that
have been mentioned in the introduction are satisfied, \rm{i.e.} (i) the vacuum
$\phi=0$ is supersymmetric and (ii) the fermions have a well defined (finite)
mass
matrix at the vacuum $\phi=0$ of the theory. (The
 fibre metric $k$ is chosen as $k={\rm diag}(1,\dots,1)$.)  To prove that the
scalar potential
$V_N(\phi^1,\dots,\phi^N)$ of the $A_N^{(1)}$ Affine Toda field theory can be
written as a sum of squares in such a way that both properties (i) and (ii)
above
are satisfied, we first observe that the scalar potential of
$A_2^{(1)}$ Affine Toda field theory  is written as
$$\eqalign{
V_2(\phi^1, \phi^2)&={m^2\over 4 \beta^2}\big[ (e^{\beta{\phi^1\over
2}} -e^{\beta{\phi^2\over 2}})^2
\cr &
+(e^{-\beta{\phi^1+\phi^2\over2}}-1)^2+8\sinh^2(\beta{\phi^1+\phi^2\over4})\big]}
\eqn\atoneaa
$$
We, then, proceed inductively in the integer $N$.  For this we assume that the
scalar potential of $A_{N-1}^{(1)}$ Affine Toda field theory can be written as
a
sum of squares that satisfy the properties (i) and (ii) above. Next we observe
that the scalar potential of the $A_N^{(1)}$ Affine Toda field theory can be
written as
$$\eqalign{
V_N(\phi^1, \dots, \phi^N)&={m^2\over 4\beta^2}\big[(e^{\beta{\phi^1\over
2}} -e^{\beta{\phi^2\over 2}})^2+\sum^N_{i\geq 3}(e^{\beta {\phi^i\over2}}-1)^2
\cr &
+(e^{-{\beta\over2}{\sum_{i=1}^N\phi^i}}-1)^2\big]+ 2
V_{N-1}({\phi^1+\phi^2\over2}, {\phi^3\over2}, \dots, {\phi^N\over2})\ . }
\eqn\atoneab
$$
This completes the induction in $N$ because from \atoneab\  the scalar
potential of $A_N^{(1)}$ Affine Toda field theory is expressed in terms of sums
of
squares and the scalar potential of  the $A_{N-1}^{(1)}$ Affine Toda field
theory.
It is easy to verify that the scalar potential of  $A_N^{(1)}$
Affine Toda field theory written as in \atoneab\ satisfies both the above (i)
and
(ii) properties. It is also straightforward to compute the section $s$ from
\atoneab\ and then use the general formalism summarised in section 2 to write
the
action of the (1,0)-supersymmetric extension of the $A_N^{(1)}$ Affine Toda
field
theory.

The above (1,0)-supersymmetric extension of the $A_N^{(1)}$ Affine  Toda field
theory is not unique.  To illustrate this, we first observe that the scalar
potential of $A_2^{(1)}$ Affine Toda field theory can also be written as
$$\eqalign{
V_2(\phi^1, \phi^2)&= {m^2\over 4 \beta^2}\big[ 2 e^{-\beta{\phi^2\over2}}
\sinh^2(\beta{2\phi^1+\phi^2\over 4})+2 e^{-\beta{\phi^1\over2}}
\sinh^2(\beta{2\phi^2+\phi^1\over 4})
\cr &
+{1\over2} \big(
[e^{\beta{\phi^1\over2}}-1]^2+[e^{\beta{\phi^2\over2}}-1]^2\big)+
2\sinh^2(\beta{\phi^1\over4})+2\sinh^2(\beta{\phi^2\over4})\big]\ .}
\eqn\atoneb
$$
We then can show that the scalar potential of $A_N^{(1)}$ Affine Toda field
theory can be written as
$$\eqalign{
V_N(\phi^1, \dots, \phi^N)&={m^2\over 4\beta^2}\big[{4\over N}\sum_{i=1}^N
e^{-\beta\sum_{j\not=i}{\phi^j\over2}}\sinh^2
(\beta{\phi^i+\sum^N_{j=1}\phi^j\over 4})
\cr &
+{N-1\over N}\sum_{i=1}^N\big (e^{\beta{\phi^i\over2}}-1\big)^2\big]
\cr &
+{2\over N} \sum^N_{i=1} V_{N-1}({\phi^1\over 2}, \dots,\hat{\phi^i\over
2},\dots
{\phi^N\over 2})\ ,}
\eqn\atonec
$$
where the `hat' over a component of the field $\phi$ implies that this
component
does not appear in the expression of the associated scalar potential. Using
induction  as in the previous case and the eqns \atoneb\ and \atonec , we can
show
that the scalar potential of $A_N^{(1)}$ Affine Toda field theory is expressed
in
terms of sums of squares and both the properties (i) and (ii) above are
satisfied.  This is a different (1,0)-supersymmetric extension of the
$A_N^{(1)}$
Affine Toda field theory from the previous one because the number of
$\psi_-$ fermions is different in these two theories. The (1,0) supersymmetry
charge is
$$
S_+=\int dx [<\partial_\pp\phi,\lambda_+>-{i\over 2} m s_a
\psi^a_-]\ ,
\eqn\atfive
$$
where $s$ is the section that one derives from either \atoneab\ or \atonec.

To give an explicit example of a supersymmetric Affine Toda field
theory, we consider the case $N=2$. Using \atoneaa, we find that the section
$s$
is
$$
\{s_a\}={1\over \beta} \{ e^{\beta {\phi^1\over2}}-e^{\beta {\phi^2\over2}},
e^{-\beta {\phi^1+\phi^2\over2}}-1, \sqrt{8} \sinh(\beta
{\phi^1+\phi^2\over4})\}\ .
\eqn\atfivea
$$
So  $s$ is a three component vector and to construct an
(1,0)-supersymmetric action with bosonic part given in \atone\ it is clear that
one
should introduce three $\{\psi^a_-; a=1,2,3\}$ chiral fermions in addition to
the two $\lambda^i_+\equiv\a^i(\lambda_+)$ chiral fermions. The action of the
supersymmetric $A_2^{(1)}$ Affine Toda field theory is
$$\eqalign{
I=&\int d^2x\,
\big[<\partial_\pp\phi,\partial_=\phi>+i<\lambda_+,\partial_=\lambda_+>-i\delta_{ab}
\psi_-^a\partial_\pp \psi_-^b
\cr &
+m\sum^2_{i=1} \lambda^i_+\partial_i s^a\psi_-^a-{m^2\over
4\beta^2}\big(\sum^2_{i=0}  e^{\beta a_i(\phi)}-3\big)\big]\ .}
\eqn\atthree
$$
The fermion mass matrix $M$ is a $2\times 3$ matrix and the
masses of the fermions at the vacuum $\phi=0$ are
$$
M\equiv m\{\partial_is_a\}(\phi=0)= m\pmatrix{{1\over2}& -{1\over2}&
{1\over \sqrt{2}}\cr
-{1\over2}& -{1\over2}& {1\over \sqrt{2}}}\ .
\eqn\atfour
$$
This matrix has rank two and so there are two Majorana fermions constructed
from
the six pairs $(\lambda_+, \phi_-)$ of Majorana-Weyl fermions that have
non-zero
mass at the supersymmetric vacuum. The  (1,0) supersymmetry charge is
computed by substituting, $s$, of eqn. \atfivea\ into \atfive.

In the case of the second (1,0)-supersymmetric extension of Affine Toda field
theory above, the section $s$ for $A_2^{(1)}$ model derived from eqn. \atoneb\
is
$$
\eqalign{
s\equiv \{s_a\}=& {\sqrt{2 }\over \beta}\{  e^{-\beta {\phi^2\over 4}}
\sinh(\beta
{2\phi^1+\phi^2\over 4})\ , e^{-\beta{ \phi^1\over 4}} \sinh(\beta
{2\phi^2+\phi^1\over 4})\ ,
\cr
&{1\over 2}( e^{\beta {\phi^1\over 2}}-1)\ ,{1\over 2}(
e^{\beta {\phi^2\over 2}}-1)\ , \sinh (\beta{\phi^1\over 4})\
,\sinh (\beta{\phi^2\over 4}) \}\ . }
\eqn\attwo
$$
So $s$ is a six component vector and to construct a (1,0)-supersymmetric action
with bosonic part given in \atone\ it is clear that one should introduce six
$\{\psi^a_-; a=1,\dots, 6\}$ chiral fermions in addition to the two
$\lambda^i_+\equiv\a^i(\lambda_+)$ chiral fermions.  The action of this
(1,0)-supersymmetric $A_2^{(1)}$ Affine Toda field theory is as \atthree\
except
that the section $s$ is given in \attwo.  The fermion mass matrix $M$ is a
$2\times 6$ matrix and the masses of the fermions at the vacuum $\phi=0$ are
$$
M\equiv m\{\partial_is_a\}(\phi=0)= {m\over \sqrt {2}}\pmatrix{1& {1\over2}&
{1\over2}& 0& {1\over 2}& 0\cr
{1\over2}& 1& 0& {1\over2}& 0&  {1\over 2}}\ .
\eqn\atfour
$$
This matrix has rank two and so there are two Majorana fermions constructed
from
the twelve pairs $(\lambda_+, \phi_-)$ of Majorana-Weyl fermions that have
non-zero
mass at the supersymmetric vacuum.    The (1,0) supersymmetry charge is
computed
by substituting, $s$, of eqn. \attwo\ into \atfive.

Next we can easily verify that in the special case of sinh-Gordon theory
($A_1^{(1)}$ Affine Toda field theory),  it is possible to further extend the
model to a (1,1)-supersymmetric one.  The scalar potential in this
case is
$$
V(\phi)={m^2\over 4\beta^2} \big(\cosh(\beta\phi)-1\big)={m^2\over 2\beta^2}
\sinh^2{\beta \phi\over 2}
\eqn\atsix
$$
and $s$ can be chosen to be simply
$$
s={\sqrt {2}\over \beta} \sinh(\beta{\phi\over2})\ .
\eqn\atseven
$$
This section $s$ can be written as the derivative of a function $f$,
$s={d\over d\phi} f$, where
$$
f=2{\sqrt {2}\over \beta^2 }\cosh(\beta{\phi\over2})\ .
\eqn\ateight
$$
The (1,0) and (0,1) supersymmetry charges of the theory are
$$
S_+=\int dx [{1\over2}\partial_\pp\phi \lambda_+ - im{1\over \beta \sqrt {2}}
\sinh(\beta {\phi\over2}) \psi_-]\ ,
\eqn\atnine
$$
and
$$
S_-=\int dx [i {1\over \sqrt {2}} \partial_=\phi \psi_-
 + m {1\over 2 \beta } \sinh(\beta {\phi\over2})\  \lambda_+]\ ,
\eqn\atten
$$
respectively, and their Poisson bracket algebra is isomorphic to the (1,1)
supersymmetry algebra section 2.

\chapter {Concluding Remarks}

We have constructed (1,0)-supersymmetric extensions of Conformal and
$A_N^{(1)}$ Affine Toda field theories using the results known for
finding supersymmetric extensions of bosonic two-dimensional sigma models. In
the
case of Conformal Toda field theory, the (1,0)-supersymmetric theory can be
further extended to a (1,1) one.  In the case of
$A_N^{(1)}$, $N\geq 2$, Affine Toda field theory, we have found that the theory
admits a (1,0)-supersymmetric extension such that the vacuum of the theory is
supersymmetric and the mass matrix of the fermions is finite at this vacuum.
We
have also demonstrated that this (1,0)-supersymmetric extension is not unique.
It
seems likely that supersymmetric extensions similar to those presented here
exist
for the rest of the Affine Toda field theories as well and for other integrable
systems like for example the Conformal Affine Toda field theory, the
non-Abelian
Toda field theory and the Moser-Calogero systems.

It is straightforward to extend the above results to the associated
one-dimensional systems, i.e. to the one-dimensional Conformal and
Affine Toda theories.  In one dimension there are no chiral fermions but one
can
still introduce the $\lambda$ and $\psi$ fermions and perform a similar
construction for the associated supersymmetric theory.  The corresponding
supersymmetric systems will have $N=1$ one-dimensional supersymmetry.

Another question of interest is that of the integrability properties of the
(1,0)-supersym\-metric Affine Toda field theories
presented in this paper.  The algebraic methods used to construct integrable
supersymmetric Toda theories in refs.[\asea, \aseb] do not seem to be
applicable
in this case because they rely on the properties of off-shell (1,1) superfields
and
such (1,1) off-shell superfield formulation for the (1,0)-supersymmetric Affine
Toda field theory constructed here does not exist. Nevertheless, one can
construct a large number of classical solutions of (1,0)-supersymmetric
$A_N^{(1)}$ Affine Toda field theory by observing  that the
configurations for which the fermions vanish and the boson fields satisfy the
field equations of the associated bosonic theory
solve the field equations of the  (1,0)-supersymmetric theory.  A consequence
of this is that for imaginary coupling constant $\beta$, the
(1,0)-supersymmetric
Affine Toda field theory will admit soliton solutions [\tjh, \hou] which
interpolate between the different classical vacua of the theory. We remark
however
that not all classical vacua of the bosonic Affine Toda field theory with
imaginary
coupling are supersymmetric.

\vskip 0.5cm

\noindent{\bf Acknowledgments:}  I am supported by a University Research
Fellowship from the Royal Society.

\refout

\bye